\documentclass[aps, prb,superscriptaddress,showpacs,noshowkeys,reprint,floatfix]{revtex4-1}
\pdfpageattr {/Group << /S /Transparency /I true /CS /DeviceRGB>>}
\usepackage[pdftex]{graphicx}
\usepackage{amsmath}
\usepackage[utf8]{inputenc}
\usepackage[T1]{fontenc}
\usepackage[finnish,english]{babel}
\usepackage{verbatim}
\usepackage{float}


\begin{document}

\title{The Structure and Local Variations of the Graphene Moiré on Ir(111)}
\author{Sampsa K. H\"{a}m\"{a}l\"{a}inen }
\affiliation{Department of Applied Physics, Aalto University School of Science, P.O.Box 11100, 00076 Aalto, Finland}
\author{Mark P. Boneschanscher}
\author{Peter H. Jacobse}
\author{Ingmar Swart}
\affiliation{Condensed Matter and Interfaces, Debye Institute for Nanomaterials Science,
Utrecht University, PO Box 80000, 3508 TA Utrecht, the Netherlands}
\author{Katariina Pussi}
\affiliation{Department of Mathematics and Physics, Lappeenranta University of Technology, P.O. Box 20 FIN-53851 Lappeenranta, Finland}
\author{Wolfgang Moritz}
\affiliation{Department of Earth and Environmental Sciences, Crystallography Section, LMU, Theresienstr. 41, D-80333 München, Germany}
\author{Jouko Lahtinen}
\author{Peter Liljeroth}
\author{Jani Sainio}
\email[]{jani.sainio@aalto.fi}
\affiliation{Department of Applied Physics, Aalto University School of Science, P.O.Box 11100, 00076 Aalto, Finland}

\date{\today}

\begin{abstract}
We have studied the incommensurate moiré structure of epitaxial graphene grown on iridium(111) by dynamic low energy electron diffraction [LEED-I(V)] and non-contact atomic force microscopy (AFM) with a CO terminated tip. Our LEED-I(V) results yield the average positions of all the atoms in the surface unit cell and are in qualitative agreement with the structure obtained from density functional theory (DFT). The AFM experiments reveal local variations of the moiré structure: the corrugation varies smoothly over several moiré unit cells between 42 and 56 pm. We attribute these variations to the varying registry between the moiré symmetry sites and the underlying substrate. We also observe isolated outliers, where the moiré top sites can be offset by an additional 10 pm. This study demonstrates that AFM imaging can be used to directly yield the local surface topography with pm accuracy even on incommensurate 2D structures with varying chemical reactivity.
\end{abstract}

\maketitle


\section{Introduction}

Determination of the geometric surface structure down to the atomic level is crucial in understanding the correlation between electronic and geometric structure. The total structure determination is particularly challenging in the case of 2D-overlayers, which are very prominent due to the rise of graphene (G) \cite{Geim:2007uq} and related materials such as hexagonal boron nitride\cite{Xue:2011fk} and silicene.\cite{PhysRevLett.108.155501,PhysRevLett.109.056804}

These atomically thin materials typically exhibit a moir\'e pattern arising from the lattice mismatch with the substrate,\cite{Wintterlin20091841,busse2011graphene,Xue:2011fk} which has been shown to cause a significant change in the electronic structure in the case of graphene.\cite{PhysRevLett.102.056808,NaturePhys.8.382,Nature.497.594,Nature.497.598} Weakly interacting overlayers are generally not commensurate with the substrate, which might result in a longer length scale modulation of the moir\'e pattern and variation of the electronic properties. Determination of the surface structure by standard tools such as dynamic low-energy electron diffraction [LEED-I(V)], scanning tunneling microscopy (STM) and atomic force microscopy (AFM) is complicated due to the large size of the moir\'e pattern and the resulting variations in the local density of states and chemical reactivity.\cite{B801785A,ANIE:ANIE200700234,Corso09012004,PhysRevB.76.075429,PhysRevLett.104.136102,PhysRevB.88.125433} These difficulties have been illustrated by numerous examples on epitaxial graphene on metal single crystal substrates.\cite{Wintterlin20091841,growth_of_graphene,PhysRevB.80.245411,doi:10.1021/j100070a027,Klink1995250,Land1992261} The graphene-substrate interaction depends on the metal, leading to a variation in the electronic and topographic structure of the moir\'e,\cite{Wintterlin20091841} and reactivity of the graphene layer.\cite{PhysRevB.83.081415,1367-2630-10-4-043033,doi:10.1021/nn3040155,PhysRevLett.97.215501,PhysRevLett.105.236101,PhysRevLett.100.056807}

The structure of the moiré on the weakly bound systems has proven to be particularly difficult to study experimentally. In STM the contrast of the moiré on G/Ir(111) inverts as a function of bias and tip termination\cite{1367-2630-10-4-043033, PhysRevB.83.081415} and the results of AFM experiments depend on the tip reactivity and the tip-sample distance.\cite{voloshina2012electronic, PhysRevB.83.081415, doi:10.1021/nn3040155}
On molecular systems, chemical functionalization of the AFM tip apex with a CO molecule and working in the repulsive force regime have become the standard way to obtain atomic scale information.\cite{Gross28082009,doi:10.1021/nn3040155,PhysRevLett.108.086101,NatCommun.4.2023,PhysRevLett.111.106103} However, all these measurements with a CO tip in the repulsive force regime have been done at constant height without AFM feedback and hence, do not yield direct information on the actual topography of the surface.

Here, we use both LEED-I(V) and scanning probe measurements to unravel the structure of the G/Ir(111) surface. The average adsorption height, registry and moiré structure are obtained from LEED-I(V) measurements where the atomic positions are described by Fourier components.\cite{PhysRevLett.104.136102} Finally, we use AFM in feedback mode with a CO terminated tip to probe local variations in the moiré structure.

\section{Methods}

The LEED measurements and graphene growth were conducted in a single ultra-high-vacuum (UHV) system with a base pressure $\approx$~$10^{-10}$~mbar. The Ir(111) single crystal was cleaned by repeated cycles of sputtering with 1.5 kV Ar$^+$ ions and subsequently annealing to 1350~K. A full monolayer of graphene was grown on the clean Ir(111) surface by chemical vapor deposition (CVD) from ethylene at 1350 K as described in Ref.  \onlinecite{growth_of_graphene}. Prior to the LEED measurements, the quality of the sample was checked with a RHK UHV-750 variable temperature STM.
 \begin{figure}[t]
\includegraphics[width=8.6cm]{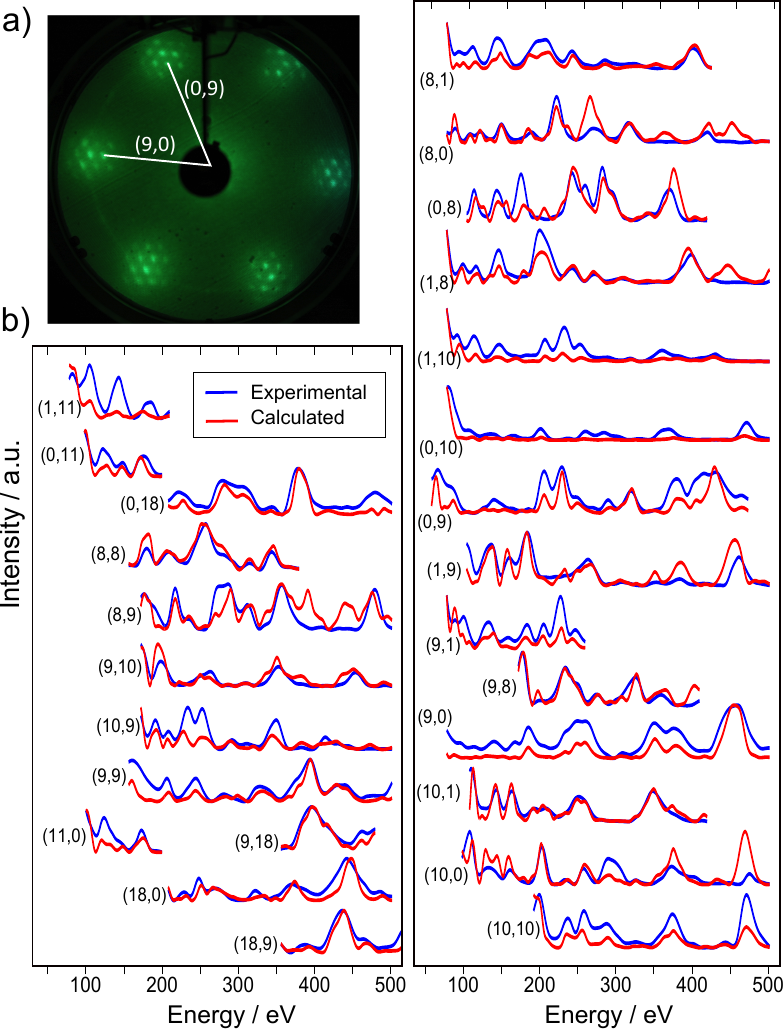}
\caption{(Color online) (a) LEED pattern of graphene on iridium (78~eV) showing the moiré spots. The illustrated indexed spots correspond to the first order substrate spots. (b) Comparison of the experimentally measured (blue) and calculated (red) I(V)-curves for the best fit (7886 eV, $R_\mathrm{P}$ = 0.39).}
\label{IVs}
\end{figure}

Princeton Research Instruments rear view LEED optics were used to measure the LEED patterns with the sample held at room temperature. The diffraction patterns were recorded in 2 eV steps from the phosphorous screen with a computer controlled Nikon D70s camera using a flat gradation curve and a single exposure setting for all the images. Due to the small spacing of the moiré diffraction spots, the background of the adjacent spots could result in false peaks in the extracted I(V) spectra. This was avoided by taking cross-sections over the spots and subtracting a linear background, similarly as in Ref \onlinecite{PhysRevLett.104.136102}. The sum of the RGB channels of the color images were used as the intensity signal.

The AFM measurements were done on a separate UHV system with an Omicron LT-STM/AFM operated at 5~K using a qPlus tuning fork with an oscillation amplitude of 85~pm. A sub monolayer of graphene was grown on the iridium crystal in order to leave clean iridium for tip preparation.  This was done by depositing a monolayer of ethylene on the clean Ir(111) and subsequently heating to 1500~K for 30 seconds.\cite{growth_of_graphene} AFM experiments were carried out in the constant frequency shift mode with a CO-terminated tip.\cite{PhysRevLett.80.2004}
To prepare the tips, CO was deposited on the graphene/Ir(111) sample at 5 K by back filling the vacuum outside the cryostat to $10^{-9}$ mbar and opening a shutter on the radiation shield for 10 s. Sometimes this already resulted in a CO terminated tip. If this was not the case, a CO molecule was picked up from bare Ir by controlled contacts by the tip. The presence of a CO molecule at the tip apex results in an inversion of the moiré contrast in STM feedback mode at low bias, which gives a simple qualitative indication of the tip termination (metal vs. a CO molecule). This effect has been confirmed in our previous work where the tip was prepared on a Cu(111) surface and then used for STM and AFM on graphene on Ir(111).\cite{doi:10.1021/nn3040155}

The LEED structure analysis was performed for a Ir(111)-(9$\times$9)-graphene-(10$\times$10) structure involving 200 C atoms per unit cell and 243 Ir atoms from the three relaxed Ir layers. In reality the system is incommensurate but the error made in the graphene lattice constant by forcing a commensurate structure is well under 1\%.\cite{busse2011graphene,1367-2630-10-4-043033} The dataset consisted of 26 beams presented in Figure \ref{IVs} with an energy range between 40 and 520 eV. The total energy range of the set was 7886 eV.

LEED calculations were restricted to models with p3m1 symmetry which was experimentally observed. The beam set neglect method was used.\cite{PhysRevLett.51.778} Convergence was checked by comparison with a full calculation for one model. The phase shifts were calculated
from a superposition of atomic potentials using optimized muffin-tin radii.\cite{PhysRevB.68.125405} Eleven phase shifts were used. A least-squares scheme was used to optimize the structural and thermal parameters in the graphene and top three substrate layers.\cite{PhysRevB.46.15438} To reduce the number of free parameters, the modulation was described by Fourier coefficients limited to the third order. Higher-order Fourier coefficients did not improve the final agreement. Lateral shifts were considered for the top graphene layer only, but no clear improvement to the agreement was gained. Overall 12 independent Fourier components for lateral and vertical modulations in the graphene layer and in three substrate layers were optimized together with 4 interlayer distances.

\section{Results and Discussion}

\begin{figure*}[t!]
\includegraphics[width=0.98\linewidth]{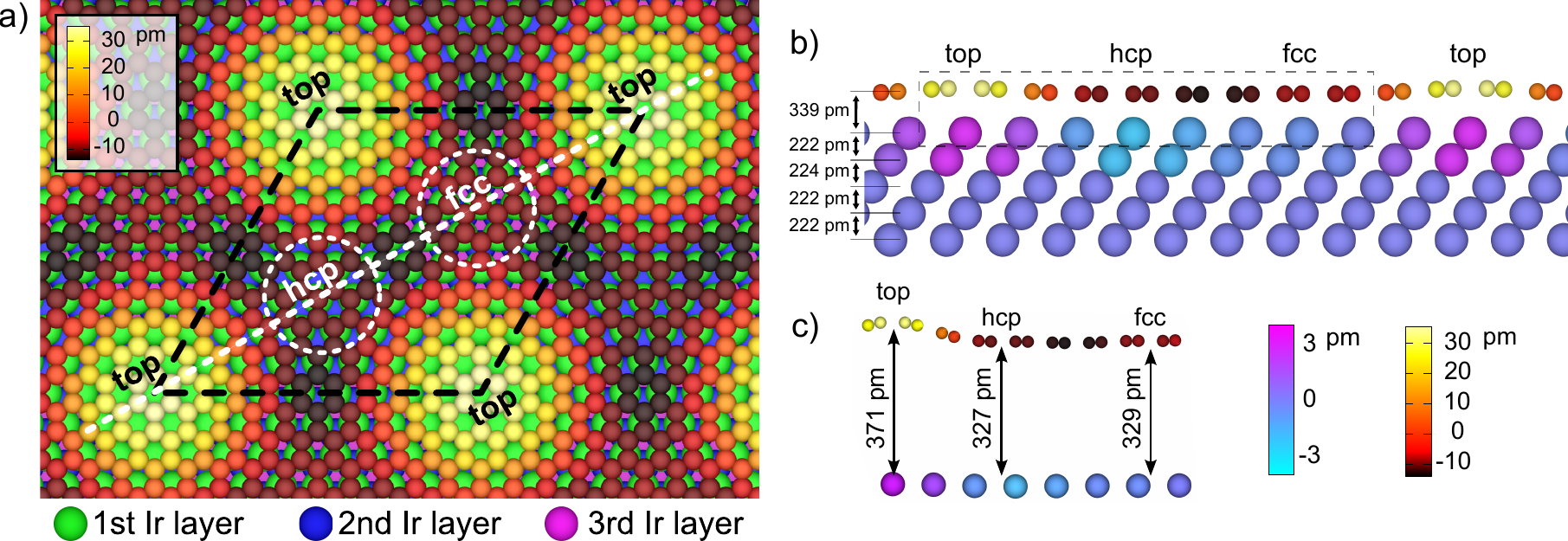}
\caption{(Color online) (a) The topographic structure of the moiré unit cell obtained from LEED-I(V). (b) Cross section through the moiré unit cell of the LEED-I(V) model along the white dashed line marked in (a). The color scales of the z-positions of the atoms are relative to the mean height of the layer, given on the left side of the image. (c) Magnification of the area marked with the dashed box in (b). The vertical scale in (c) is magnified 5-fold to better illustrate the shape of the graphene layer.}
\label{section}
\end{figure*}

We will first discuss the LEED-I(V) results before moving on the AFM data. The structure of the moiré unit cell obtained from the LEED-I(V) calculations is presented in Figure \ref{section} and the I(V) curves in Figure \ref{IVs}. The structure largely agrees with a previous vdW-DFT study.\cite{busse2011graphene} Similar to other graphene/metal systems, the graphene Ir distance is largest where the center of the carbon ring is directly above an Ir atom (top site) (Fig. \ref{section}a). The smallest graphene-Ir separation is found in the bridge site region between the hcp and fcc sites (see Fig. \ref{section}a for explanation of abbreviations). This is in contrast to the vdW-DFT results where the bridge site region is higher than either of the hollow site regions. The R-factor of the calculated LEED-I(V) data is not very sensitive to the Fourier component causing this small (2~pm) dip around the bridge site and and hence, this is likely to be an artifact in the model (see below for AFM results). Not taking into account the dip on the bridge site the overall corrugation of the graphene layer in the model is (43$\pm$9)~pm. This is slightly higher, although within the error margins, than the value predicted by vdW-DFT (35~pm).\cite{busse2011graphene}

The graphene-Ir separations in the hcp and fcc areas match exactly the values given by vdW-DFT (327~pm and 329~pm, respectively).\cite{busse2011graphene} It is worth noting that the optimization for the LEED structure was started from a completely flat layer of graphene not to introduce any bias in the structure. The larger corrugation in our results compared to vdW-DFT is caused by the height of the top site where the graphene-Ir separation from our LEED-I(V) analysis is 371~pm compared to 362~pm given by vdW-DFT.  The mean height of the graphene is (339$\pm$3)~pm, which is in excellent agreement with both XSW and vdW-DFT.\cite{busse2011graphene} The first two Ir layers in the best fit LEED structure are also slightly corrugated in phase with the graphene layer. This corrugation is however within the limits of error of the LEED-I(V) calculation. No significant stretching of the C-C bonds was observed in the LEED model, which is also inline with the DFT results.\cite{private}

In addition to the LEED-I(V) analysis, we have studied the atomic scale corrugation of the moiré with AFM using a CO terminated tip. A CO molecule on the end of the AFM tip has been shown to be chemically inert on the graphene/Ir(111) system.\cite{doi:10.1021/nn3040155} When scanning very close to the surface, Pauli repulsion is the dominant force between the tip and carbon atoms.\cite{Gross28082009} Thus the AFM experiments yield the actual topography of the graphene surface not distorted by local variations in chemical reactivity or local density of states.

\begin{figure}[b!]
\includegraphics[width=0.8\linewidth]{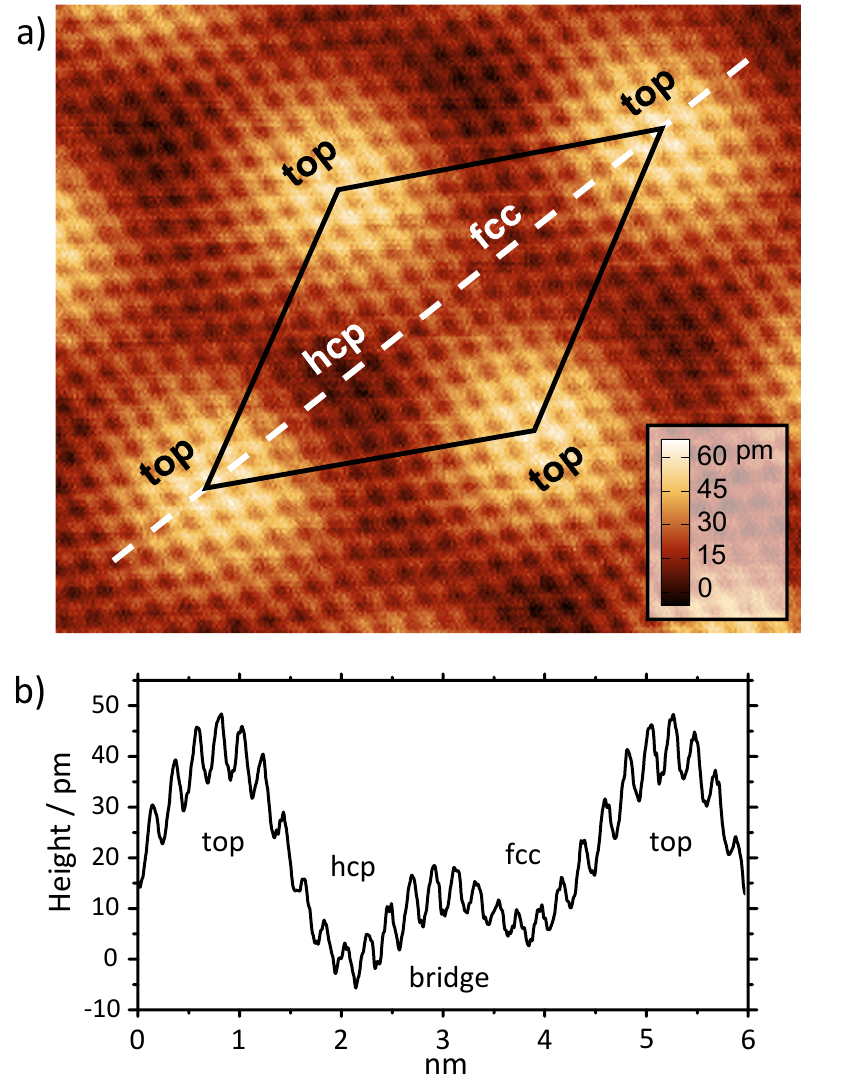}
\caption{(Color online) (a) Constant frequency shift AFM image of the graphene moiré ($\Delta f$ = 0 Hz, bias = 0 V). (b) AFM line profile over the moiré unit cell marked with a white dashed line in panel (a).}
\label{afm}
\end{figure}

Figure \ref{afm}a shows an AFM image acquired with a CO terminated tip. Qualitatively the structure is very similar to that obtained from LEED-I(V) and predicted by vdW-DFT. We can relate the bright hills to the on top sites of the moiré unit cell by comparing to STM images acquired immediately before and after the AFM image. We assign the lower of the fcc and hcp sites to the hcp site, inline with LEED-I(V) measurements and DFT calculations. In contrast to the LEED-I(V) structure, the AFM images (cross-section in Fig. \ref{afm}b) show that the bridge site is higher than the fcc and hcp sites, which is in agreement with the vdW-DFT results.

\begin{figure}[pt]
\includegraphics[width=0.95\linewidth]{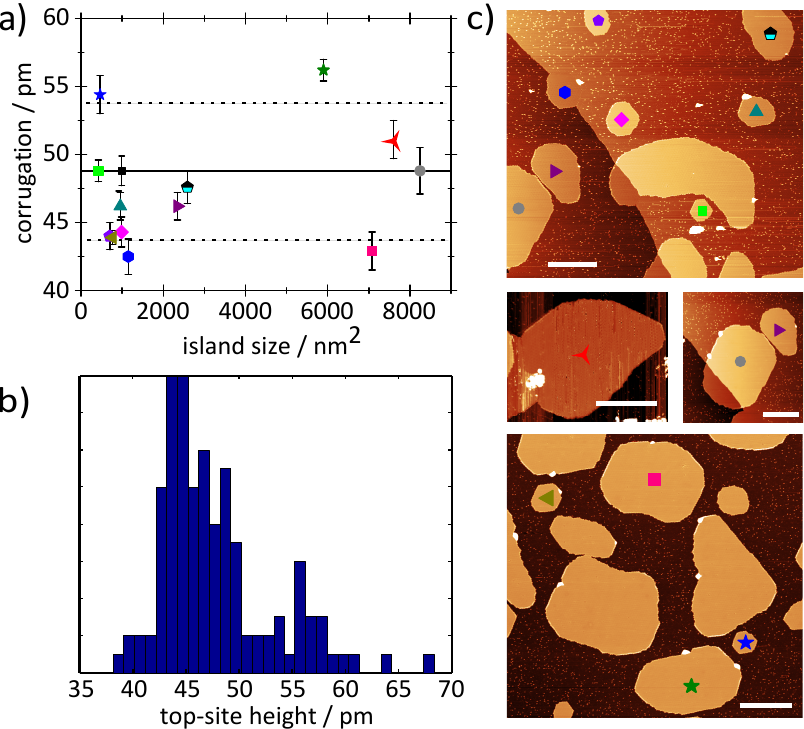}
\caption{(Color online) (a) The moiré corrugation as a function of graphene island size excluding the outliers. The solid line is the average of the set and the dotted lines the standard deviation. (b) Distribution of all the individual top site heights extracted from the AFM images (including outliers). (c) Overview STM scans of the islands in the plot in panel (a). The scale bar is 50 nm in all images.}
\label{islands}
\end{figure}

Based on XSW measurements, it has been suggested that the moiré corrugation of graphene on Ir is not constant, but changes as a function of graphene coverage.\cite{busse2011graphene} As a local probe AFM can be used to study the order and corrugation of individual graphene islands as a function of their size and environment. We have imaged 14 islands of various sizes and shapes \ref{islands}, with some of them flowing over or growing  from steps and others lying freely on an iridium terrace.

The obtained moiré corrugations as a function of island size are plotted in Fig. \ref{islands}a. The AFM measurements were all conducted with a CO tip with a detuning set point of 0~Hz, which corresponds to a repulsive interaction between the CO molecule on the tip and graphene. The tip was characterized before each image by measuring the frequency shift and current as a function of tip-sample distance. The corrugations were extracted from atomically resolved 8$\times$8 nm$^2$ images by comparing the height of the top most atoms from each top site to the lowest atoms of the neighboring three hcp sites.

\begin{figure}[pt]
\includegraphics[width=0.98\linewidth]{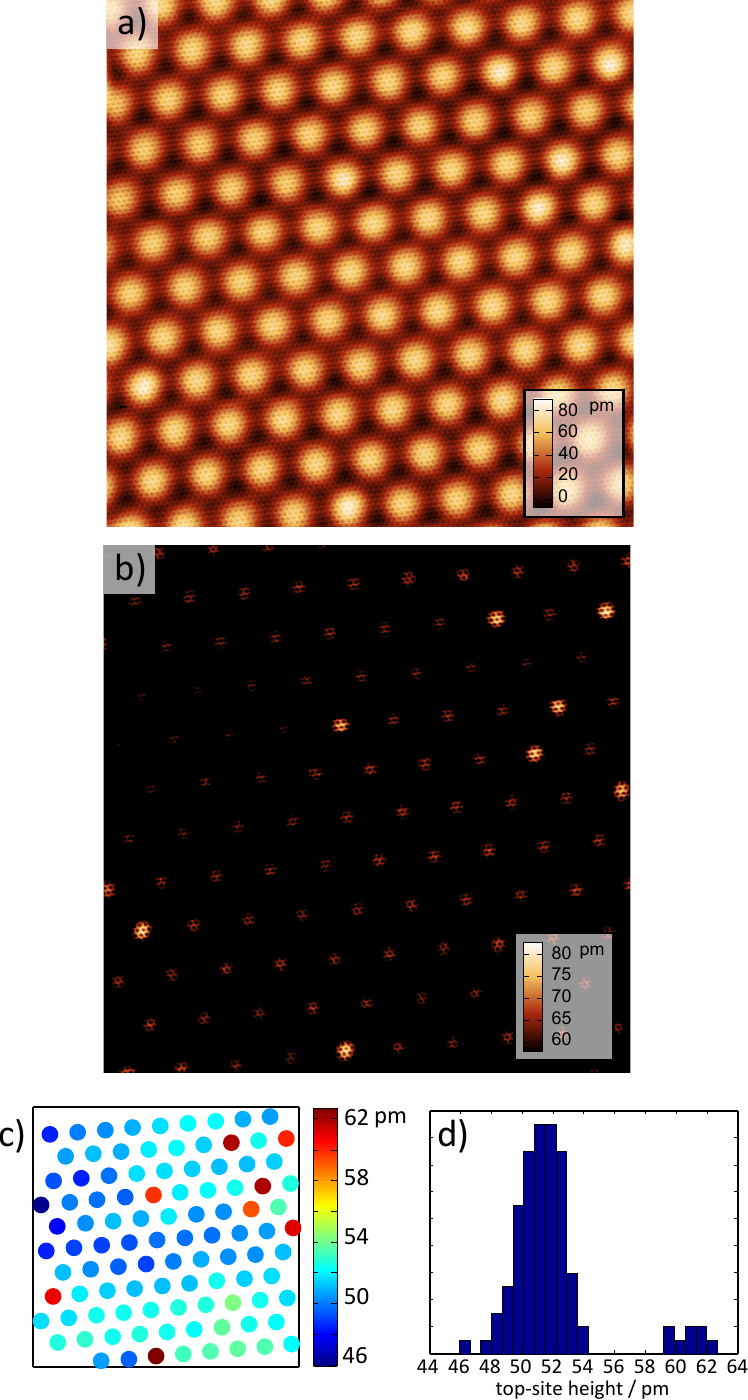}
\caption{(Color online) (a) Constant frequency shift AFM image over a larger area ($\Delta f$ = 0 Hz, bias = 0 V, 24$\times$24 nm$^2$).  (b) Same AFM image as in panel (a) but with the contrast adjusted to the top sites. (c) Heights of the top sites with respect to the neighboring hcp sites. (d) Distribution of the top site heights in panel (c).}
\label{fig4}
\end{figure}

The data shown in Fig. \ref{islands}a shows no definite trend between the island size and moiré corrugation in the studied size range (420$\dots$8200~nm$^2$). What is surprising though is the spread of the measured corrugations (42$\dots$56~pm). Neither the moiré rotation, environment or island aspect ratio had any correlation with the measured corrugations. Based on the data shown in Fig. \ref{islands}, we obtain an estimate of the total corrugation of (47$\pm$5)~pm which is inline with the LEED-I(V) model. The height difference between the fcc and the hcp sites is (5$\pm$2)~pm.

While analyzing the corrugations we noticed that some of the images had outliers where the top site of one moiré unit cell was much higher than the rest (these cells were excluded from Fig. \ref{islands}a and from the average values above). To study the outliers in detail we imaged a much larger area (24$\times$24~nm$^2$) with atomic resolution from one of the islands (Fig. \ref{fig4}a). When limiting the contrast of the image to the top sites (Fig. \ref{fig4}b), it is easy to distinguish the higher outliers which are randomly distributed around the scanned area. The atomic contrast on the outliers varies, which indicates that they do not correspond to a specific graphene-substrate registry. Figure \ref{fig4}c shows the height of each top site with respect to the neighboring hcp sites. As can be seen from the extracted heights the outliers are all roughly the same height. This is even more evident when plotting the heights in a histogram (Fig. \ref{fig4}d) where the outliers show up as a distinct peak some 10 pm higher then the rest of the top sites. A similar peak is visible in the histogram in Fig. \ref{islands}b with all the measured moiré heights.

In addition to the outliers, the top site heights in Fig. \ref{fig4}c also exhibit a smooth variation over several moiré unit cells. This variation within one island is of the same magnitude as the differences between the different islands in Fig. \ref{islands}. There is a difference between the smooth long range variation and the outliers. The outliers are moiré unit cells where the top site is lifted higher, whereas in the longer range fluctuations, the heights of both the top and the hcp site vary.


We will now discuss possible sources of the observed variations in the moiré structure. The unit cell of the moiré structure of graphene on iridium has been shown to be incommensurate with respect to both the graphene and Ir lattices.\cite{PhysRevB.86.235439,1367-2630-10-4-043033} This means that the graphene carbon rings are not exactly on the symmetry sites which are used to describe the structure (top, hcp, fcc) but change slightly from one moiré to the next. In a very simple model, this should produce a repeating 2nd order moiré structure, where the symmetry sites are closer to the ideal case in some regions than in others. This could affect the local interaction between graphene and iridium leading to variation in the adsorption heights of the different areas. This is a possible source for the long range variation in the observed moiré corrugation as it would most likely affect both the strongly and weakly bound\cite{busse2011graphene} hollow and top sites of the moiré. The 2nd order moiré is not expected to be rigid and the structure would be likely to exhibit fluctuations, inline with the variations shown in Fig. \ref{fig4}.

Spot profile analysis low energy electron diffraction has shown that graphene locks to the iridium substrate already at high temperatures, which upon cooling induces strain in the graphene lattice due to the mismatch in thermal expansion coefficients.\cite{doi:10.1021/nl203530t,PhysRevLett.106.135501,PhysRevB.88.165406} This locking most likely happens at the edges of the graphene island which strongly interact with the iridium substrate.\cite{PhysRevLett.103.166101,doi:10.1021/nn3040155,ADMA:ADMA201204539} The strain is partially relieved in large islands by local delamination into wrinkles.\cite{1367-2630-11-11-113056} The islands we studied are so small that no wrinkling was observed in any of them. The outliers could however be a way to relieve some of the strain before complete local delamination. The top sites are weakly bound by vdW forces\cite{busse2011graphene} and hence would be the first sites to accommodate the strain. As these outliers do not correspond to a specific graphene-substrate registry, it is unlikely that they are related to the 2nd order moiré effect discussed above.
\vspace{0.62cm}
\section{Conclusion}

In conclusion, we present the first experimental investigation on the local variations in the moir\'e structure of an incommensurate graphene layer. We attribute these variations to a second order moir\'e structure where the exact registry of the moiré symmetry sites changes from one unit cell to the next. Our dynamic low-energy electron diffraction experiments yield the average structure and registry of the moiré pattern of epitaxial graphene on Ir(111). Comparing this structure with AFM experiments shows that AFM imaging can be used to directly yield the local surface topography with pm accuracy on 2D structures. This type of information is likely to be important in detailed understanding of the electronic properties of weakly interacting incommensurate 2D structures, such as graphene on hexagonal boron nitride.

\begin{acknowledgments}
We thank Ari P. Seitsonen and Carsten Busse for discussions. This research was supported by the the Finnish Academy of Science and Letters, the Academy of Finland (Centre of Excellence in Low Temperature Quantum Phenomena and Devices No. 250280 and project No. 218186), the European Research Council (ERC-2011-StG No. 278698 "PRECISE-NANO"), and FOM [Control over Functional Nanoparticle Solids (FNPS)]. We acknowledge the computational resources provided by Finland's IT Center for Science (CSC). This research made use of the Aalto Nanomicroscopy Center (Aalto NMC) facilities.
\end{acknowledgments}

\end{document}